\pdfoutput=1
\documentclass[letterpaper]{article}
\usepackage{times}
\usepackage{helvet}
\usepackage{courier}
\usepackage{graphicx}
\usepackage{subfig}

\usepackage{graphicx}
\usepackage{amsmath}
\usepackage{amssymb}
\usepackage{mathabx}
\usepackage{algorithm2e}
\usepackage[backref=page]{hyperref}

\newtheorem{theorem}{Theorem}
\newtheorem{lemma}{Lemma}
\newtheorem{defn}[theorem]{Definition}
\newcommand{\qed}{\hfill \ensuremath{\Box}}
\newenvironment{proof}[1][Proof.]{\begin{trivlist}
\item[\hskip \labelsep {\bfseries #1}]}{\hfill\qed\end{trivlist}}

\newcommand{\BE}{\begin{enumerate}} \newcommand{\EE}{\end{enumerate}}
\newcommand{\BEQN}{\begin{eqnarray}} \newcommand{\EEQN}{\end{eqnarray}}
\newcommand{\BEQNN}{\begin{eqnarray*}} \newcommand{\EEQNN}{\end{eqnarray*}}
\newcommand{\BEQ}{\begin{equation}} \newcommand{\EEQ}{\end{equation}}

\newtheorem{algo}{Algorithm}
\newcommand{\BA}{\begin{algo}} \newcommand{\EA}{\end{algo}}

\newcommand{\BI}{\begin{itemize}} \newcommand{\EI}{\end{itemize}}

\newtheorem{theo}{Theorem}

\newcommand{\BT}{\begin{theo}} \newcommand{\ET}{\end{theo}}
\newcommand{\BL}{\begin{lemma}} \newcommand{\EL}{\end{lemma}}
\newcommand{\BCM}{\begin{clm}} \newcommand{\ECM}{\end{clm}}

\newcommand{\BPF}{\begin{proof}}
\newcommand{\EPF}{\end{proof}}

\newenvironment{proofof}[1]{\noindent{\bf Proof of {#1}.~}}{\endprf}
\newcommand{\BPFOF}{\begin{proofof}} \newcommand {\EPFOF}{\end{proofof}}

\def\set#1{\{#1\}}

\begin{document}
\title{Approximation and Heuristic Algorithms for Probabilistic Physical Search on General Graphs}
\author{Noam Hazon \\ noamh@ariel.ac.il \\ \and Mira Gonen \\ mirag@ariel.ac.il \\ \and Max Kleb \\ maxkleb@gmail.com \\ \\
		Department of Computer Science and Mathematics\\
		Ariel University, Israel}
\maketitle
\begin{abstract}
We consider an agent seeking to obtain an item, potentially available at different locations in a physical environment. The traveling costs between locations are known in advance, but there is only probabilistic knowledge regarding the possible prices of the item at any given location.
Given such a setting, the problem is to find a plan that maximizes the probability of acquiring the good while minimizing both travel and purchase costs.
Sample applications include agents in search-and-rescue or exploration missions, e.g., a rover on Mars seeking to mine a specific mineral. These probabilistic physical search problems have been previously studied, but we present the first approximation and heuristic algorithms for solving such problems on general graphs. We establish an interesting connection between these problems and classical graph-search problems, which led us to provide the approximation algorithms and hardness of approximation results for our settings. We further suggest several heuristics for practical use, and demonstrate their effectiveness with simulation on real graph structure and synthetic graphs.
\end{abstract}

\section{Introduction}
An autonomous intelligent agent often needs to explore its environment and choose among different available alternatives. In many physical environments the exploration is costly, and the agent also faces uncertainty regarding the price of the possible alternatives.
For example, consider a traveling purchaser seeking to obtain an item~\cite{Ramesh81TPP}.
While there may be prior knowledge regarding candidate stores (e.g., based on search history), the actual price at any given site may only be determined upon reaching the site.
In another domain, consider a Rover robot seeking to mine a certain mineral on the face of Mars.
While there may be prior knowledge regarding candidate mining sites (e.g., based on satellite images), the actual cost associated with the mining at any given location, e.g., in terms of battery consumption, may depend on the exact conditions at each site (e.g., soil type, terrain, etc.), and hence are fully known only upon reaching the site.

These scenarios are referred to as probabilistic physical search problems, since there is a prior probabilistic knowledge regarding the price of the possible alternatives at each site, and traveling for the purpose of observing a price typically entails a cost. Furthermore, exploration and obtaining the item results in the expenditure of the same type of resource. The purchaser's  money is used not only to obtain the item but also for traveling from one potential store to another; the robot's battery is used not only for mining the mineral but also for traveling from one potential location to another. Thus, the agent needs to carefully plan its exploration and balance its use of the available budget between the exploration cost and the purchasing cost.

This paper focuses on the development of efficient exploration strategies for probabilistic physical search problems on graphs. The analysis of such problems was initiated by~\cite{aumann:2008,hazon2013physical}, who showed that it is (computationally) hard to find the optimal solution on general graphs. Accordingly, they provide a thorough analysis of physical search problems on one-dimensional path graphs, both for single and multi-agent settings
However, many real-world physical environments may only be represented by two-dimensional graphs. For example, the Mars rover can freely move directly from any possible mining location to another (with an associated travel cost), while in path graphs the robot is restricted to move only to the two adjacent neighbors of its current location. Our work thus handles probabilistic physical search problems on general graphs.
To the best of our knowledge, our is the first to do this.

We consider two variants of the problem. The first variant,
coined \textit{Max-Probability}, considers an agent that is given an initial budget for the task (which
it cannot exceed) and needs to act in a way that maximizes
the probability it will complete its task (e.g., reach at least one opportunity with a budget large enough to successfully
buy the product). In the second variant, coined \textit{Min-Budget}, we are required to guarantee
some pre-determined success probability, and the goal
is to minimize the initial budget necessary in order
to achieve the said success probability.

Since previous work showed that probabilistic physical search problems are hard on general graphs, we either need to consider approximations with guaranteed bounds or heuristics with practical running time. We do both. We first establish an interesting connection between \textit{Max-Probability} and the \textit{Deadline-TSP} problems~\cite{Bansal:04}, and as a result we are able to provide an $O(\log n)$ approximation for the former, based on an $O(\log n)$ approximation for the latter, with the only requirement that the probabilities are not too small. We then show a $5+\epsilon$ approximation, for every $\epsilon>0$, for the special case of \textit{Min-Budget} with equal purchasing costs, equal single probabilities, and a hardness of approximation within a ratio of $1.003553$ for the general \textit{Min-Budget} problem.
We then consider heuristics for practical use. We suggest two families of heuristics, linear-time and exponential-time heuristics. We evaluate the performance of our heuristics through simulations on graphs extracted from a real network and on synthetic graphs.
We found that our no-backtrack branch-and-bound algorithm is able to efficiently solve very large instances while producing solutions that are very close to optimal, even though it has a theoretical exponential worst-case running time. Our ant-colony based heuristic, which has a linear worst-case running time, does not lag much behind.

\section{Related Work}
Models of search processes with prior probabilistic knowledge have been studied extensively in the economic literature~\cite{Weitzman1979,McMillan94Rothschild}. However, these economic-based search models assume that the cost associated with observing a given opportunity is stationary, i.e., does not change along the search process. In our settings the agent is operated in a physical environment. That is, the distances to other sites depends on the agent's position, and thus when the agent travels to explore other sites the cost of traveling to other sites changes.

Changing search costs has been considered in the computer science domain traditionally in the contexts of the
\textit{Prize-Collecting Traveling Salesman} problem~\cite{Balas:89} and its variants.
These problems, while related, fundamentally differ from our model in that the traveling budget and the prizes in these models are distinct, with different “currencies”. Thus, expending the travel budget does not, and cannot affect the prize collected at a node. In our work, in contrast, traveling and buying use the same resource (e.g. battery power).
The \textit{Deadline-TSP}~\cite{Bansal:04} problem, which is a generalization of the \textit{Orienteering} problem~\cite{Tsiligirides:84}, is much more relevant to our settings, and in the next section we establish a connection between the \textit{Max-Probability} and the \textit{Deadline-TSP} problems.

The work most related to ours is the work of~\cite{hazon2013physical}, who introduced the probabilistic physical search problems and provided a comprehensive analysis of the problems on one-dimensional path graphs, both for single~\cite{aumann:2008} and multi-agent settings~\cite{hazon:2009}.
Recently,~\cite{brown:15} presented an MILP formulation and a branch-and-bound optimal algorithm for general graphs, which work only if the graph is complete.
In a different paper,~\cite{brown15kagents} investigate the minimal number of agents required to solve \textit{Max-Probability} and \textit{Min-Budget} problems on a path and in a 2-dimensional Euclidean space.

From a broader perspective, our search problems relate to planning, scheduling and path planning with uncertainty.
We refer the reader to~\cite{hazon2013physical} for a comprehensive overview of relevant works and how they relate to probabilistic physical search.

\section{Preliminaries}
We are given a graph $G=(V,E)$, whose vertices represent the sites where an item is available (i.e. stores) and the edges represent the connections between the sites. We are also given a weight function $w$ on the edges $w:E\rightarrow R^+$, which determines the travel costs between any two sites. W.l.o.g. (without loss of generality) we assume that the agent's initial location is at one of the sites, denoted by $v_1$, and the item cannot be obtained at this site. The cost of obtaining the item at each site $v \in V$ is a random variable $C_v$ with an associated probability mass function $p_v(c_i)$ for $1\le i\le k$, which gives the probability that obtaining the item will cost $c_i$ at site $v$. For ease of notation, we assume that all sites have $k$ cost values (if a site $v$ has fewer cost values then we can add arbitrary cost values $c_i$ with $p_v(c_i) = 0$). Hence, with a  probability of $1-\sum_{i=1}^k{p_v(c_i)}$ the item cannot be obtained at a given site $v$. Note that the actual cost at any site is only revealed once the agent reaches the site.

The total cost for the agent includes both the traveling cost and the cost of obtaining the item.
The agent travels along a path $\mathcal{P} = \langle v_1,\ldots,v_{\ell}\rangle$ where $(v_i,v_{i+1})\in E$, which is an ordered multiset of vertices.
That is, $v_i \in \mathcal{P}$ represents the $i$-th vertex of the path, and thus it is possible that for $i \neq j, v_i = v_j$.
Notice that we allow the agent to visit a site $v$ multiple times if needed. However, the probability $p_v(c_i)$ of obtaining the item at cost $c_i$ will be counted only once - specifically, the first time the agent reaches $v$ with a remaining budget of at least $c_i$. 
The cost of traveling a path $\mathcal{P} = \langle v_1,\ldots,v_{\ell}\rangle$ is $\sum_{i=1}^{\ell-1}{w(v_i,v_{i+1})}$, hereafter  denoted $w_\mathcal{P}$.
Given these inputs, the goal is to find a path that maximizes
the probability of obtaining the item, while minimizing the necessary budget.
The standard approach in such multi-criterion optimization problems is to optimize one of the objectives while bounding the
other. In our case, we get two concrete problem formulations (following~\cite{hazon2013physical}):
\BE
\item Max-Probability: given a total initial budget $B$, maximize the probability of obtaining the item.
\item Min-Budget: given a success probability $p_{succ}$, minimize the budget needed to guarantee the item will be obtained with a probability of at least $p_{succ}$.
\EE

%
\section{Max-Probability}
In this section we provide an $O(\log n)$ approximation algorithm for the \emph{Max-probability} problem, when the probabilities are not too small.
We first consider a restricted case of the \emph{Max-Probability} problem, and then we show how to extend our analysis to the general case.
Our algorithm is built on the approximation algorithm of~\cite{Bansal:04} for the \emph{Deadline-TSP} problem that they defined
as follows:
\begin{defn}
Given a weighted graph $G=(V,E)$ on $n$ nodes, with a start node $r$, a prize function $\pi:V\rightarrow Z^+$, deadlines $D:V\rightarrow Z^+$, and a length function $\ell:E\rightarrow Z^+$, find a path starting at $r$ that maximizes the total prize, where a path starting at $r$ collects the prize $\pi(v)$ at node $v$ if it reaches $v$ before $D(v)$.
\end{defn}
If $\mathcal{P}$ is a path found by an algorithm for the \emph{Deadline-TSP},let $V(\mathcal{P})$ denote the set of nodes visited by $\mathcal{P}$ before their deadline.

\subsection{The case of single probabilities} \label{sec:single_prob}
We begin by providing an approximation algorithm for the case where $k=1$. That is, in each site either the item can be obtained at a given cost with a given probability, or not available at all. We abuse the notation and use $c_v$ to denote the (single) cost of the item at site $v$, and $p_v$ to denote the probability of obtaining the item there.

Our goal is to maximize the probability of success using a given budget. That is, the probability that the agent will be able to succeed in at least one site, and that the total cost is at most the given budget B.
We may also phrase our objective as minimizing the failure probability.
Formally, we would like to find a path  $\mathcal{P} =  \langle v_1,\ldots,v_{\ell}\rangle$ with a set of vertices $V(\mathcal{P}) \subseteq \mathcal{P}$ such that:
\begin{itemize}
 \item $w_\mathcal{P} \le B$.
 \item For all $v_j \in \mathcal{P} \setminus \{v_1\}$, if $v_j\in V(\mathcal{P})$ then $B-\sum_{i=1}^{j-1}{w(v_i,v_{i+1})} \ge c_v$ and if $v_j\notin V(\mathcal{P})$ then $B-\sum_{i=1}^{j-1}{w(v_i,v_{i+1})}< c_v$.
 \item For all $v_i=v_j\in \mathcal{P}$,  if $i<j$ and $v_i\in V(\mathcal{P})$, it holds that  $v_j\notin V(\mathcal{P})$.
 \item $\prod_{v\in V(\mathcal{P})} (1-p_v)$ is minimal.
\end{itemize}
However,
\BEQN \label{eq:obj}
\small
&&\arg\min_{\mathcal{P}}\{\prod_{v\in V(\mathcal{P})} (1-p_v)\} =\nonumber\\
&&\arg\min_{\mathcal{P}}  \{\log(\prod_{v\in V(\mathcal{P})} (1-p_v))\} =\nonumber\\
&&\arg\max_{\mathcal{P}}  \{-\log(\prod_{v\in V(\mathcal{P})} (1-p_v))\} =\nonumber\\
&&\arg\max_{\mathcal{P}} \{\sum_{v\in V(\mathcal{P})} -\log(1-p_v)\}
\EEQN
Therefore it is suffice to find a path such that $\sum_{v\in V(\mathcal{P})}{-\log(1-p_v)}$ is maximal.
Since we represent our objective as an optimization over summation we can convert every instance of our problem into an instance of the \textit{Deadline-TSP} problem and run the approximation algorithm of~\cite{Bansal:04}. This is not straightforward, since the \textit{Deadline-TSP} is defined over prizes from $Z^+$ but due to our conversion the prizes will not be necessarily integers
\footnote{Although the lengths of edges in the \textit{Deadline-TSP} problem are integers, and in the \textit{Max-Probability} problem they are not necessarily so, we note that they do not play any role in the optimization process.}.
Lemma~\ref{lem:scale} shows that we can overcome this challenge, if we bound the size of the prizes by any small constant.
\BL\label{lem:scale}
If there is an $r$-approximation algorithm for the \emph{Deadline-TSP} problem where the prizes are integers, 
then there is an $O(r)$-approximation algorithm for the \emph{Deadline-TSP} problem where the prizes are not necessarily integers, but there is a lower bound of $1/c$ on these prizes, for any constant $c$ larger than $0$.
\EL
\BPF
Given an instance with prizes $\pi(v)$ for every vertex $v$, that are not necessarily integers, let $\pi'(v) = \lfloor\pi(v)\rfloor$ if $\pi(v)\ge 1$, and $1$ otherwise.
Let $OPT$ be the value of the optimal solution with the original prizes and $\cal P_{OPT}$ an optimal path. $OPT'$ and $\cal P_{OPT'}$ are similarly defined with the scaled prizes $\pi'(v)$.
Then,
\BEQN \label{eq:scale}
OPT'&=&\sum_{v\in V(\cal P_{OPT'})}{\pi'(v)}\ge
\sum_{v\in V(\cal P_{OPT})}{\pi'(v)}\nonumber\\&\ge&\sum_{v\in V(\cal P_{OPT})}{\left(\frac{\pi(v)}{2}\right)}=
\frac{1}{2}OPT.
\EEQN
The last inequality is true since $\pi'(v)\ge 1$, thus by rounding we lose at most a factor of $2$.
%
Now, suppose we have an $r$-approximation algorithm for the \emph{Deadline-TSP} problem that uses rounded integer prizes. Let $Alg'$ be the total prize collected by this algorithm and ${\cal P}_{Alg'}$ the path returned by this algorithm. Clearly, $Alg'\ge \frac{1}{r}\cdot OPT'$. If we use the path ${\cal P}_{Alg'}$ with the non-rounded prizes we will collect a different total prize, denoted $Alg$. If $c\le 1$ then no prize has been rounded up, and according to Equation~\ref{eq:scale},
\BEQN
Alg&\ge&  Alg'\ge \frac{1}{r}\cdot OPT'\ge \frac{1}{2r}\cdot OPT.
\EEQN
If $c>1$ then in the rounding process each prize of a node in $V({\cal P}_{Alg'})$ is increased by at most $1-1/c$,
 so it holds that
\BEQN
Alg&\ge&  Alg'-|V({\cal P}_{Alg'})|\cdot(1-1/c)\nonumber\\&\ge& Alg'-Alg'\cdot(1-1/c)= Alg'\cdot1/c\nonumber\\
&\ge&  \frac{1}{r}\cdot \frac{1}{c}\cdot OPT'\ge \frac{1}{2r\cdot c}\cdot OPT.
\EEQN
Therefore, the lemma immediately follows.
\EPF

We are now ready to present the details of our conversion and show how it guarantees an approximation ratio of $O(\log{n})$.
The idea is that the probabilities correspond to prizes and the costs correspond to deadlines, but the challenge here is to keep the same approximation ratio.
\BT
The \emph{Max-Probability} problem for the case of single probabilities can be approximated within a ratio of $O(\log n)$, for any instance of the problem for which it holds that $p_v\ge 1/c$  for each probability $p_v$, where $c$ is any constant larger than 1.
\ET
\BPF
Given an instance of the \emph{Max-Probability} in which $p\ge 1/c$ for every probability $p$, we construct an instance of \emph{Deadline-TSP} as follows. Each site of the \emph{Max-Probability} is a node of \emph{Deadline-TSP}, $r=v_1$, and the weight function $w$ becomes the length function $\ell$. For each site $v$, we set the prize at the corresponding node, $\pi(v)$,  to $-\log(1-p_v)$ and the deadline $D(v)$ to $B-c_v$.
Note that in the  \emph{Deadline-TSP} problem when starting at $r$ the time is $0$, and it increases when traveling the path. However, in the \emph{Max-Probability} problem the budget, $B$, is maximal when starting at $v_1$, and it decreases when traveling the path. Therefore, by setting $D(v)$ to $B-c_v$ we can consider the budget already spent instead of the remaining budget. The budget already spent is $0$ when starting at $v_1$, and increases when traveling the path. When reaching a node $v$ in which $D(v) = B-c_v$, if the budget already spent is at most $D(v)$ the item can be bought at $v$ since the remaining budget is at least $c_v$.

Now, we apply the approximation algorithm of~\cite{Bansal:04} to the \emph{Deadline-TSP} problem with the instance described above, and we use the path returned as a solution for \emph{Max-Probability}. 
Obviously, $p\ge 1/c$ implies that $-\log(1-p)\ge -\log(1-1/c)$. For every constant $c>1$ a constant $c'\neq 0$ exists such that $-\log(1-1/c)\ge 1/c'$. (For instance take $c'=-\frac{1}{\log(1-1/c)}$). Therefore, by Lemma~\ref{lem:scale} we can use the approximation algorithm of~\cite{Bansal:04} even for non-integer prizes and lose only a factor of $c'$ in the approximation ratio. 

It remains to show that the approximation ratio is $O(\log n)$. Let $p^*$ be the optimal probability of the instance of \emph{Max-Probability} problem.
Let $T^*$ be the optimal total prize of the \emph{Deadline-TSP} problem on the converted instance, 
and let ${\mathcal P}^*$ be a path returned by an optimal algorithm for the problem.
Similarly, let 
$T^{apx}$ be the total prize collected by the approximate algorithm for the \emph{Deadline-TSP} problem on the converted instance, and let 
${\mathcal P}^{apx}$  be the path returned by the approximation algorithm. Finally, let 
$p^{apx}$ be the probability achieved by using ${\mathcal P}^{apx}$
for solving \emph{Max-Probability}.
According to our conversion, $T^* = \sum_{v\in V(\mathcal{P}^*)}  -\log(1-p_v)$, and $T^{apx} = \sum_{v\in V(\mathcal{P}^{apx})}  -\log(1-p_v)$. 
By Equation~\eqref{eq:obj} we find that $p^* = 1 - \prod_{v\in V({\mathcal P}^*)} (1-p_v)$, and by our construction 
$p^{apx} = 1 - \prod_{v\in V({\mathcal P}^{apx})} (1-p_v)$.
According to~\cite{Bansal:04}, a constant $d$ exists such that $T^*/(d\cdot \log n) \leq T^{apx}$. To simplify notations, let $r = 1/(d\cdot\log n)$. Therefore,
\BEQNN
&&\sum_{v\in V({\mathcal P}^*)} -\log(1-p_v)\cdot r \leq \sum_{v\in V({\mathcal P}^{apx})}  -\log(1-p_v)\Rightarrow\nonumber\\
&&\log(\prod_{v\in V({\mathcal P}^*)} (1-p_v))\cdot r \geq \log(\prod_{v\in V({\mathcal P}^{apx})} (1-p_v)) \Rightarrow\nonumber\\
&&(\prod_{v\in V({\mathcal P}^*)}{(1-p_v)})^{r} \geq \prod_{v\in V({\mathcal P}^{apx})}{(1-p_v)}\nonumber\\
\EEQNN

%
%
Therefore we get:
\begin{multline*}
p^{apx} =  1-\prod_{v\in V({\mathcal P}^{apx})}{(1-p_v)} \ge  \\
1-(\prod_{v\in V({\mathcal P}^*)}{(1-p_v)})^{r} = 1-(1-p^*)^{r}.
\end{multline*}
We use the generalization of Newton's binom, namely $(1+x)^r = \sum_{j=0}^{\infty}{{r \choose j}x^j}$ for any real $r$, where ${r \choose j}=\frac{r(r-1)\ldots(r-j+1)}{j!}$ for $j>0$, and ${r \choose 0}=1$. Thus,
\begin{multline*}
1-(1-p^*)^{r} = 1-\sum_{j=0}^{\infty}{{r\choose j}(p^*)^j\cdot(-1)^j} \\
= \sum_{j=1}^{\infty}{{r\choose j}(p^*)^j\cdot(-1)^{j+1}}
\ge r\cdot p^*.
\end{multline*}
The last inequality is valid since the sum of two consecutive terms – the $(2i)$th term and the $(2i+1)$th term – is positive:
\begin{multline*}
{r\choose 2i}(p^*)^{2i}\cdot(-1)^{2i+1}+ {r\choose 2i+1}(p^*)^{2i+1}\cdot(-1)^{2i+1+1}\\
= -\frac{1}{(2i)!}\cdot r\cdot \left(r-1\right)\cdot\ldots\cdot\left(r-2i+1\right)\cdot (p^*)^{2i}\\
+ \frac{1}{(2i+1)!}\cdot r\cdot \left(r-1\right)\cdot\ldots\cdot\left(r-(2i+1)+1\right)\cdot (p^*)^{2i+1}\nonumber\\ = (p^*)^{2i}\cdot \frac{1}{(2i)!}\cdot r\cdot \underbrace{\left(r-1\right)\cdot\ldots\cdot\left(r-2i+1\right)}_{<0}\nonumber\\\cdot\underbrace{\left[-1+\frac{1}{2i+1}\cdot\left(r-2i\right)\cdot p^*\right]}_{<0}>0
\end{multline*}
Overall, we find that,
$p^{apx} \ge r\cdot p^*$, as required.
\EPF

\subsection{General Case}
We now show how to extend our results from the previous section to provide an approximation algorithm for the general case, i.e., where the number of probabilities in each site is not bounded.
W.l.o.g. assume that in each site $c_1\le c_2 \le \ldots \le c_k$. Thus, an agent that reaches a site $v$ with a reaming budget of $b$ will acquire the item with a probability of $\sum_{j=1}^i{p_v(c_j)}$ for $i$ for which $c_i\le b < c_{i+1}$ if $i<k$, and probability of $\sum_{j=1}^k{p_v(c_j)}$ if $b \ge c_k$ (namely, $i=k$ if $b\ge c_k$).

We reduce the general case to the case of single probabilities as follows.  Given an instance graph $G$ to the (general) \emph{Max-Probability} problem, we define a new graph $G'$ that will have a single probability in each site. For every vertex $v$ with a probability $p_v(c_i)$ of obtaining the item at a cost $c_i$, where $1\le i\le k$, we define new vertices $u_1,\ldots, u_k$, such that in each site $u_i$, where $2\le i\le k$, either the item can be acquired at a cost of $c_i$ with a probability $p_{u_i}=\frac{p_v(c_i)}{1-\sum_{j=1}^{i-1}{p_v(c_j)}}$, or not available at all. In $u_1$, either the item can be acquired at a cost of $c_1$ with a probability $p_{u_1}=p_v(c_1)$, or not available at all. We replace the vertex $v$ in $G'$ with $u_1$, and create edges between $u_i$ and $u_{i+1}$, for all $1 \le i \le k-1$, with associated weights of $0$.

Therefore, an agent that reaches $u_1$ in $G'$ with a remaining budget of $b$ can travel without any cost from $u_1$ to $u_k$ and back to $u_1$. For $i$ for which $c_i\le b < c_{i+1}$, or for $i=k$ if $b\ge c_k$, the probability of failure from this travel is as follows: 
\begin{multline*}
(1-p_{u_1})\cdot\ldots\cdot(1-p_{u_i})
=(1-p_v(c_1))\cdot \left(1-\frac{p_v(c_2)}{1-p_v(c_1)}\right)\cdot \\ \ldots \cdot\left(1-\frac{p_v(c_i)}{1-\sum_{j=1}^{i-1}{p_v(c_j)}}\right)
=1-\sum_{j=1}^i{p_v(c_j)},
\end{multline*}
which is the same probability of failure of an agent that reaches the corresponding site $v$ in $G$ with a reaming budget of $b$.
Therefore, the \emph{Max-Probability} problem on $G$ can be approximated using the approximation algorithm from the previous section on $G'$. 
Since $G'$ has $k\cdot n$ nodes, we conclude:
\BT
The \emph{Max-Probability} problem can be approximated within a ratio of $O(\log n+\log k)$, for any instance of the problem for which it holds that $\frac{p_v(c_i)}{1-\sum_{j=1}^{i-1}{p_v(c_j)}}\ge 1/c$ for every vertex $v$ and any cost $c_i \in C_v$, where $c$ is any constant larger than $1$. For $k=O(n)$ the \emph{Max-Probability} problem can be approximated within a ratio of $O(\log n)$, for the same instances.
\ET
%
\section{Min-Budget}
Although \emph{Min-Budget} is the dual of \emph{Max-Probability} and their decision versions are the same, it seems that \emph{Min-Budget} is much harder to approximate on general graphs. Indeed, converting the \emph{Min-Budget} problem to the dual of the \emph{Deadline-TSP} is hopeless; we have a proof that unlike the \emph{Deadline-TSP} problem, the dual of the \emph{Deadline-TSP} problem is hard to approximate within a factor of $c\cdot \log n$, for any constant $c$.  
We thus consider restricted instances. We show that the \emph{Min-Budget} problem with a specific instance of equal vertex costs and equal single probabilities can be approximated within a ratio of $5+\epsilon$ for any $\epsilon>0$. The idea is to run the $2+\epsilon$ approximation algorithm of~\cite{AroraK00} for the \emph{rooted k-MST} problem and then travel along the tree. The \emph{rooted k-MST} problem, that was shown to be NP-hard~\cite{raviSMRR94}, is as follows:
\begin{defn}
Given a graph $G=(V,E)$ on $n$ nodes with a root node $r$,
nonnegative edge weights, and a specified number $k$, find a tree of minimum weight that includes $r$, which spans at least $k$ nodes other than $r$.
\end{defn}
\BT
The \emph{Min-Budget} problem with a specific instance of equal vertex costs and equal single probabilities can be approximated within a ratio of
$5+\epsilon$ for any $\epsilon>0$.
\ET
\begin{proof}
Given an instance of \emph{Min-Budget}, we define an instance for the \emph{rooted k-MST} problem, where $r=v_1$ and,
\BEQN
k=\left\lceil\frac{\log(1-p_{succ})}{\log(1-p)}\right\rceil.
\EEQN
We then run the approximation algorithm of~\cite{AroraK00} and return the path received by traveling along the tree of~\cite{AroraK00}. The initial budget is set to $c$ plus twice the tree's cost. Obviously, the path returned by the algorithm meets all the constraints of the minimum budget problem: if the tree spans $k$ vertices then since $k\ge \frac{\log(1-p_{succ})}{\log(1-p)}$, it holds that $\log(1-p_{succ})\ge k\cdot \log(1-p)$, thus the probability that the algorithm will succeed is at least $1-(1-p)^k\ge p_{succ}$. Moreover, since we add $c$ to the budget, the item can be bought at every node in the returned path.

Now, let $M^{apx}$ be the value returned by the approximation algorithm of~\cite{AroraK00} for the \emph{rooted k-MST} problem and
let $M^*$ be the optimum value of the \emph{rooted k-MST} problem. Similarly, let $B^{apx}$ be the budget required by the above approximation algorithm for the \emph{Min-Budget} problem, and let $B^*$ be the minimal budget for the instance of the \emph{Min-Budget} problem. Thus
\BEQN
B^{apx}&=&2 \cdot M^{apx}+c\le 2\cdot(2+\epsilon)\cdot M^*+c\\ \nonumber &\le& 2\cdot(2+\epsilon)\cdot B^*+c\le (5+2\epsilon)\cdot B^*.
\EEQN
\vspace{-3pt}
\end{proof}

On the other hand, we show that the \emph{Min-Budget} problem is hard to approximate within a factor of $\alpha=1.003553$. We do so by using the hardness of approximation of the \emph{Min-Excess-Path} problem proven in~\cite{BlumCKLMM07}. The excess of a path is defines as follows: 
\begin{defn}
Given a graph $G=(V,E)$ on $n$ nodes with a root node $s$ and an end node $t$,
nonnegative edge weights, nonnegative prizes for each vertex, and a quota $Q$, the excess of a path is 
the difference between the length of an $s-t$ path that collects prizes of at least $Q$, and the length of the shortest path between
$s$ and $t$.
\end{defn}
That is, any path must spend a minimum amount of time equal to the shortest distance
between $s$ and $t$, only to reach the destination $t$; the excess of the path is the extra time it spent to gather
prizes along the way. The \emph{Min-Excess-Path} problem is to find a minimum-excess path from $s$ to $t$ collecting
prizes of at least $Q$, and~\cite{BlumCKLMM07} show that the \emph{Min-Excess-Path} problem is NP-hard to approximate to within a factor of $\beta=220/219$.
\BT
The Min-Budget problem is hard to approximate within a ratio of $\alpha=1.003553$.
\ET
\begin{proof}
We reduce the \emph{Min-Excess-Path} problem to the \emph{Min-Budget} problem. To do so we use the hard instance of the \emph{Min-Excess-Path} problem described in~\cite{BlumCKLMM07} as follows. Let $G=(V,E)$ be a complete graph on $n$ nodes, with edge weights in the set $\{1,2\}$, let $r$ be both the starting node and ending node. and let $n$ be the desired quota. W.o.l.g. assume that $n\ge 1000$. Given the above instance of the \emph{Min-Excess-Path} problem we define an instance for the \emph{Min-Budget} problem using the same graph, $v_1 = r$, and in each vertex there is a single cost of $1$ with a constant probability $p$. In addition, let $p_{succ}= 1-(1-p)^n$. Now, assume that there is an approximation algorithm for the \emph{Min-Budget} problem with an $\alpha$ ratio. We show that this implies an approximation algorithm for the \emph{Min-Excess-Path} problem with a $\beta$ ratio, thus contradicting~\cite{BlumCKLMM07}.
Let $A_B$  be an $\alpha$-approximation of the \emph{Min-Budget} problem. Then run $A_B$ on the instance defined above. Let $s$ be the ending node of the path returned by this algorithm. Then take the path returned by this algorithm, and add edge $(s,r)$ to the path, and its weight to the path's cost. The resulting cycle is a solution for the \emph{Min-Excess-Path} problem. We now analyze the approximation ratio of this algorithm. Let $A_{EX}$ be the value of the path returned by the above algorithm to the \emph{Min-Excess-Path} problem, $O_B$ the optimal value of the \emph{Min-Budget} problem for the defined instance, and $O_{EX}(u,t)$ the optimal value of the \emph{Min-Excess-Path} problem starting at $u$ and ending at $t$. Let $t$ be the ending node of an optimal \emph{Min-Budget} solution. Then,
 \BEQN
A_{EX}(r,r)&\le& A_{EX}(r,s)\le A_B-d(r,s)-1\nonumber\\&\le&\alpha\cdot O_B-d(r,s)-1\nonumber\\&\le& \alpha(d(r,t)+O_{EX}(r,t)+1)-d(r,s)-1\nonumber\\&\le_{(*)}&
\alpha(d(r,t)+O_{EX}(r,r)+1)-d(r,s)-1\nonumber\\&\le& \alpha\cdot O_{EX}(r,r)+\alpha\cdot 3-2.
\EEQN
 Notice that $O_{EX}(r,r)\ge n-2$. This is true since a path has to traverse all the nodes in order to collect enough prizes, each edge is of a length of at least 1, and the shortest path between the last vertex in which the quota is met and $r$ is at most 2. Thus we get that \BEQN
 3\cdot\alpha-2&=&\frac{3\cdot\alpha-2}{n-2}\cdot (n-2)\nonumber\\&\le& \frac{3\cdot\alpha-2}{n-2}\cdot O_{EX}(r,r)\le \beta\cdot O_{EX}(r,r).
 \EEQN
 (The last inequality is valid since $n\ge 1000$, so $1<\alpha\le \frac{\beta(n-2)+2}{n+1}$).
 This contradicts the fact that the \emph{Min-Excess-Path} problem cannot be approximated within a ratio of $\beta$~\cite{BlumCKLMM07}.

Proof of (*): Notice that $O_{EX}=\min_w\{O_{EX}(r,w)\}$, since there is a node on the cycle going through $r$ in which all the needed prizes were already collected. If node $w$ for which the minimal value is received is not $t$ (=the last vertex on the path of the above optimal value of the Min Budget) then we can take the path from $r$ to $w$ and get
\BEQN
 A_B&=&O_{EX}(r,w)+d(r,w)+1\nonumber\\&<&O_{EX}(r,t)+d(r,w)+1\nonumber\\&\le& O_{EX}(r,t)+3 = O_B-d(r,t)-1+3\nonumber\\&=&O_B-d(r,t)+2\le Q_B+1.
 \EEQN
 This implies that $A_B\le O_B$, so we can attain an optimal solution of the \emph{Min-Budget} that returns a path for which $w$ is the last vertex. Thus we can assume w.o.l.g. that $O_{EX}(r,r)=O_{EX}(r,t)$.

\vspace{-3pt}
\end{proof}

\section{Heuristics and Experimental Analysis}
Our theoretical results in the previous section led us to consider heuristics for practical use. In this section we propose several heuristics and experimentally evaluate them against the optimal solution. We concentrate on the \emph{Min-Budget} problem, and the same ideas can be used to build heuristics for the \emph{Max-Probability} problem (the implementation is even simpler). Indeed, we tested the heuristics for \emph{Max-Probability}, and even the simplest greedy heuristic almost always achieved a probability that was very close to the optimal probability.

We use the following notations. Let $\mathcal{P}$ be the path that the agent traversed hitherto, $\mathcal{P} = \langle v_1, \ldots, v_\ell \rangle$. That is, $v_\ell$ is the site where the agent is currently located. Let $\mathcal{N}_\mathcal{P}$ be the set of all neighbors of sites in $\mathcal{P}$, that is $\mathcal{N}_\mathcal{P}=\set{v|u \in \mathcal{P}, (u,v)\in E}$. If $v \in \mathcal{N}_\mathcal{P}$, let $w_{v}^\mathcal{P}$ be the total weight of the shortest path from $v_\ell$ to $v$ that uses only sites from $\mathcal{P}$.
We tested the following methods:
\begin{itemize}
	\item \textit{Optimal.} In the cases where it was computationally feasible to do so, we exhaustively evaluated every possible path with several budgets, in order to find the real optimal solution as a comparison. We implemented a branch-and-bound algorithm, that is based on the algorithm of~\cite{brown:15}, to reduce the running time. 
	\item \textit{Greedy.} Let the score of a site $v \in \mathcal{N}_\mathcal{P}$ and a cost $c_i \in C_v$ be $S(\mathcal{P},v,c_i) = \frac{\sum_{j=1}^i p_v(c_j)}{w_{v}^\mathcal{P} \cdot c_i}$.
	In each iteration, the heuristic locally chooses the next best site and the best cost there: 
	$\arg\max_{v \in \mathcal{N}_\mathcal{P}} \arg\max_{c_i \in C_v} S(\mathcal{P},v,c_i)$.
	That is, the heuristic chooses the site that has the maximal success probability to cost ratio, over all possible probabilities and costs. The heuristic then increases the initial budget so that the agent will be able to travel to the chosen site and obtain the item at the chosen cost. Note that the heuristic calculates a cost of $w_{v}^\mathcal{P} \cdot c_i$ instead of $w_{v}^\mathcal{P} + c_i$. Intuitively, it captures the ``penalty'' for exploring a distant site (with a high $w_{v}^\mathcal{P}$) that will incur a high traveling cost for returning, if needed. In addition, we experimentally tested the greedy heuristic with a denominator of $w_{v}^\mathcal{P} + c_i$, and it performed much worse. 
	\item \textit{Ant Colony Optimization (ACO).} Following the successful application of the ant colony optimization technique for producing near-optimal solutions for the Traveling Salesman problem~\cite{dorigo1996ant} and Vehicle Routing with Time Windows problem ~\cite{Bachem1996vr}, we developed an ACO version for the \emph{Min-Budget} problem, as follows. We ran $50$ iterations. In each iteration the ant chooses the next site $v \in \mathcal{N}_\mathcal{P}$ and the cost $c_i \in C_v$ with a probability of $\frac{S(\mathcal{P},v,c_i) \cdot h_v^\mathcal{P}}{\sum_{v \in \mathcal{N}_\mathcal{P}, c_i \in C_v}  S(\mathcal{P},v,c_i) \cdot h_v^\mathcal{P}}$, where $h_v^\mathcal{P}$ is the average pheromone level of the edges in the shortest path from $v_\ell$ to $v$ that uses only sites from $\mathcal{P}$. That is, the ant randomly chooses the next site and cost, where the probability of selection depends on the attractiveness of the site and the costs. The pheromone level of each edge is initially set to $1$, and after each iteration it evaporates by $0.05$. However, after finding a path $\mathcal{P}$ which is better than the current best path, the ant updates the pheromone level of each edge $(u,v)$ of the path to $w(u,v) \cdot V(\mathcal{P}) / w(\mathcal{P})$.
	\item \textit{Bounded-Length (BL).} This heuristic is a restricted version of the optimal branch-and-bound algorithm, which bounds the solution's length by two means. First, the heuristic prunes any path that is longer than the length of the best solution found so far. In addition, the heuristic does not allow the agent to traverse through an unvisited site without spending any budget there. Clearly, this restricted branch-and-bound algorithm is no longer guaranteed to be optimal. However, it is expected to run faster than the optimal algorithm since the solution's length has a major impact on the optimal algorithm's running time.  
	\item \textit{No-Backtrack (NB).} Another reason for the long running time of the optimal algorithm is the backtracking phase that checks paths with repetitions, i.e., where the agent visits the same sites more than once. This motivated us to consider a restricted version of the optimal algorithm, where in addition to bounding the length of solution (as in the BL heuristic) the algorithm does not backtrack and thus only checks paths without repetitions. Unlike the other heuristic, the NB heuristic does not necessarily find a solution for every instance.
\end{itemize}

\subsection{Experimental Design and Results}
\begin{figure}[htbp!]\centering
\includegraphics[width=0.65\columnwidth]{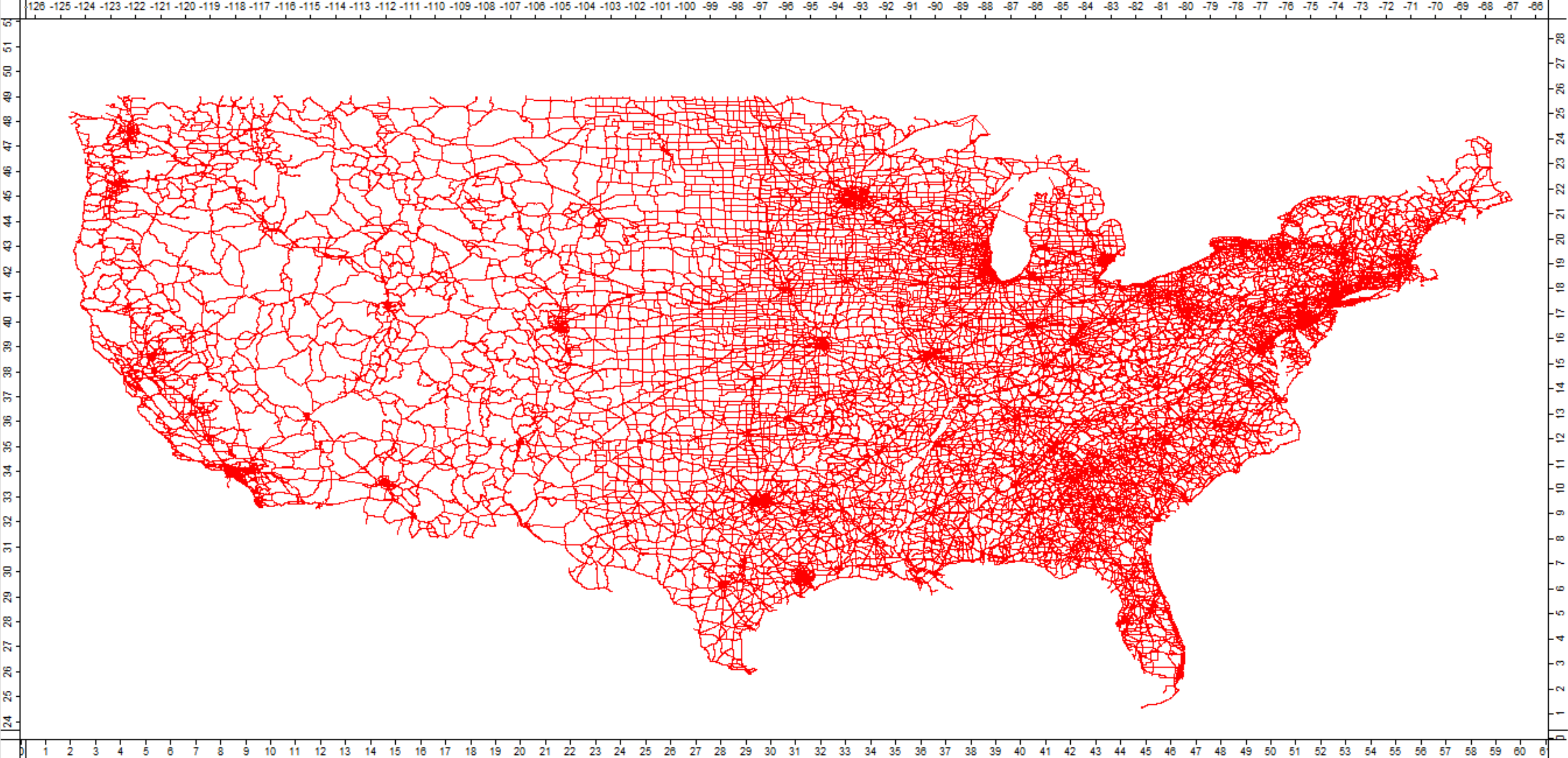}
\caption{\small The full real graph, extracted from GIS data of the highways netowrk of the USA.}\label{fig:full_graph}
\end{figure}

\begin{figure}[htbp!]\centering
\includegraphics[width=0.65\columnwidth]{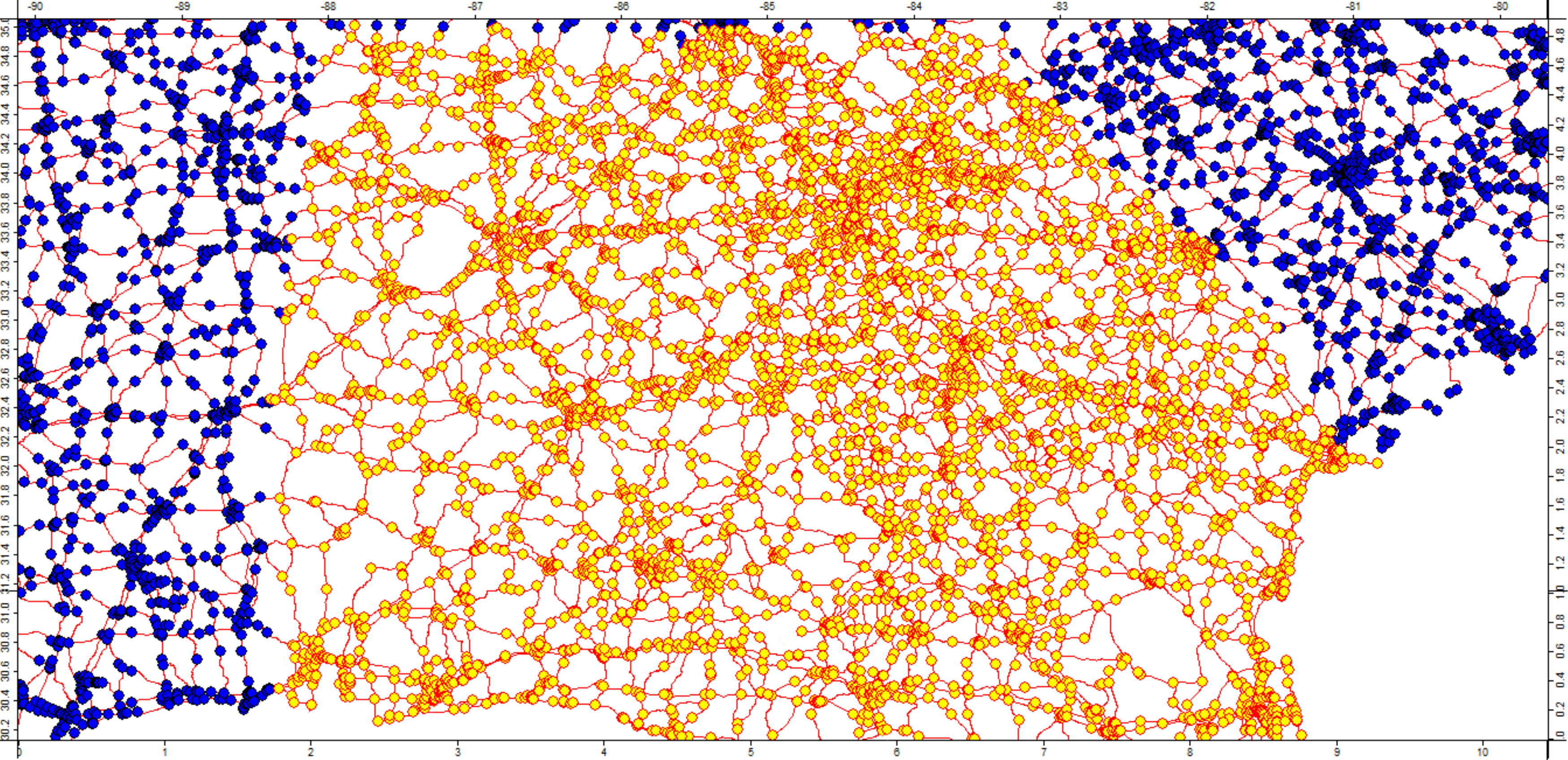}
\caption{\small An example of one of the subgraphs (the yellow nodes) that we sampled from the full netwrok.}\label{fig:sample_graph}
\end{figure}

For the empirical evaluation of our heuristics we used a real graph structure with the traveling costs set as the real distance between the vertices, which we extracted from GIS data of the highways network of the USA\footnote{http://www.mapcruzin.com/download-mexico-canada-us-transportaton-shapefile.htm}.
Since the original network is too large we sampled $40$ random sub-graphs, with an average number of $6325.5$ vertices.
An illustration of the full graph and of one of the sampled subgraphs are depicted in Figures~\ref{fig:full_graph} and \ref{fig:sample_graph}, respectively
As for the costs, for each vertex we randomly generated between $1$ and $5$ costs to obtain the item with a non-zero probability. The costs were generated using a normal distribution with an expectation of $2700$ and a standard deviation of $900$ (the costs were bounded within two standard deviations from the expectation). 

\begin{figure}[htbp!]\centering
\vspace{-10pt}
\subfloat[{\small Minimal required budegt.}]
{\includegraphics[width=0.5\columnwidth]{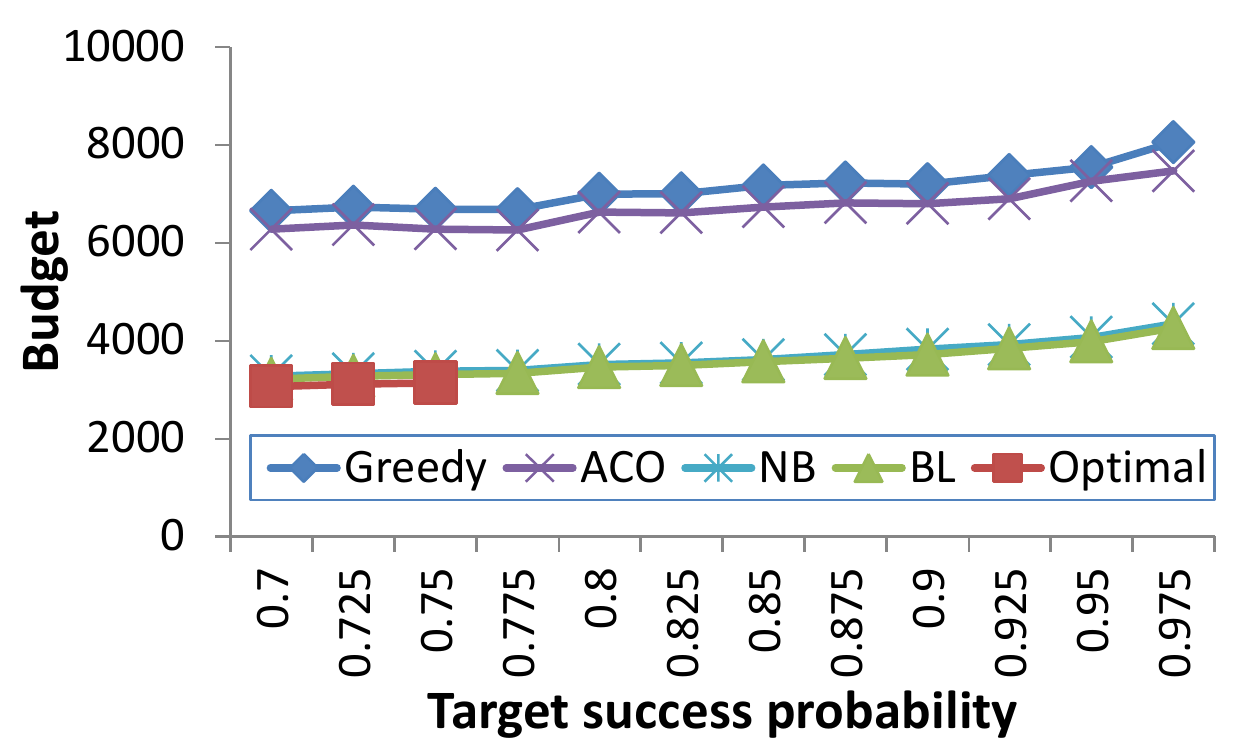}\label{subfig:prob_budget}}\hfill
\subfloat[{\small Runing time.}]
{\includegraphics[width=0.5\columnwidth]{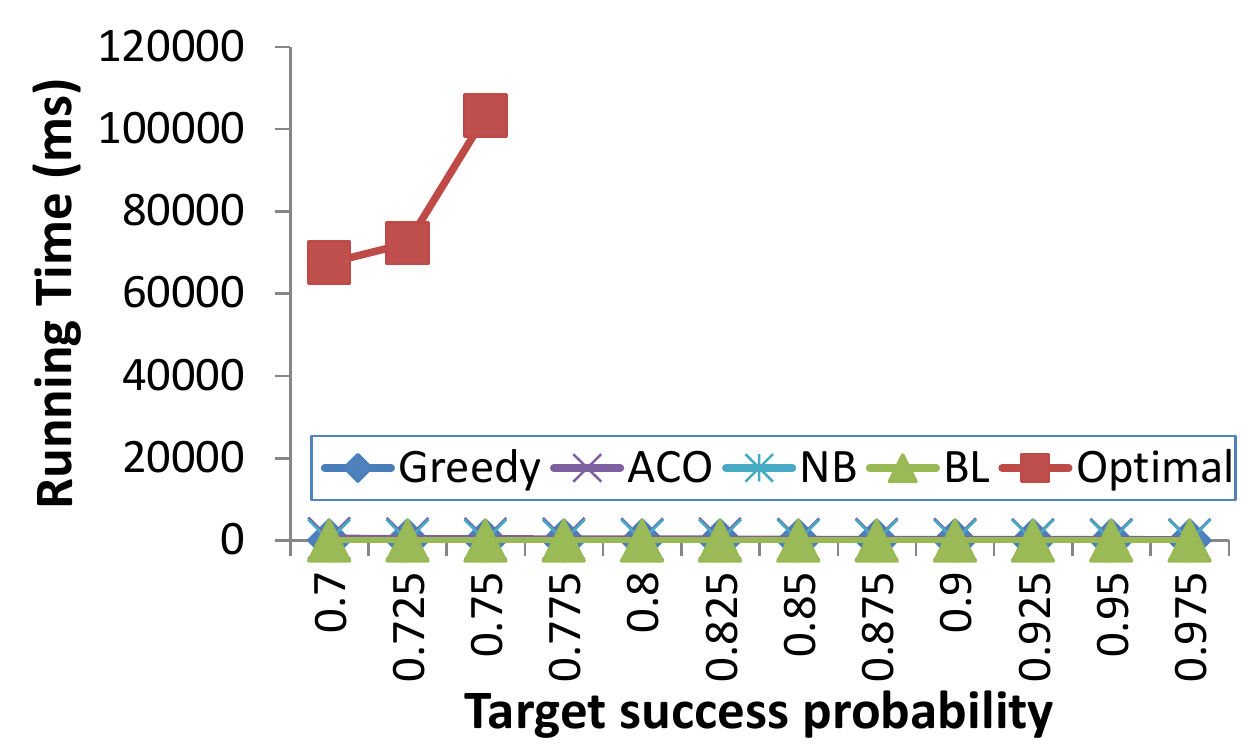}\label{subfig:prob_time}}
\vspace{-5pt}
\caption{\small Varying the target success probability, $p_{succ}$.}\label{fig:prob}
\vspace{-8pt}
\end{figure}

We began by testing the effect of the target success probability, $p_{succ}$, on the performance of our heuristics. We thus randomly generated probabilities for each vertex using a normal distribution with an expectation of $0.24$ and a standard deviation of $0.08$. We then varied $p_{succ}$ between $0.7$ and $0.975$. The results are depicted in Figure~\ref{fig:prob}, where each point is the average over the $40$ graphs.
%
%
%
%
As expected, a higher target success probability results in a higher minimal required budget and longer running time. 
However, there was no statistically significant difference between the budget required by the optimal and BL heuristic and NB was only a little behind. ACO was statistically significantly better than Greedy, but ACO still required between $72$-$91\%$ more budget than NB.
As for the running time, all of our heuristics were able to find solutions within a reasonable time, but the optimal algorithm demonstrated its anticipated exponential running time behavior in the early stages. Surprisingly, the NB heuristic was faster than almost all the heuristics (only the greedy heuristic was faster) while still producing solutions that are near-optimal. Our explanation is that the hard instances for the optimal algorithm, when there is a need to backtrack in order to find a good solution, possibly occur when the graph has many dead-ends, i.e., vertices with a degree of $1$, or with edges that are very costly. Since we use a real graph structure with real costs, such vertices are rarely found.

We also wanted to test the performance of our heuristics when we decrease the probabilities (for acquiring the item) in the vertices.
We thus randomly generated probabilities for each vertex using normal distributions, where we varied the expectation between $0.3$ and $0.09$. The standard deviation was set to a third of the expectation, and the target success probability was set to $0.90$.
%
%
\begin{figure}[htbp!]\centering
\vspace{-10pt}
\subfloat[{\small Minimal required budegt.}]
{\includegraphics[width=0.5\columnwidth]{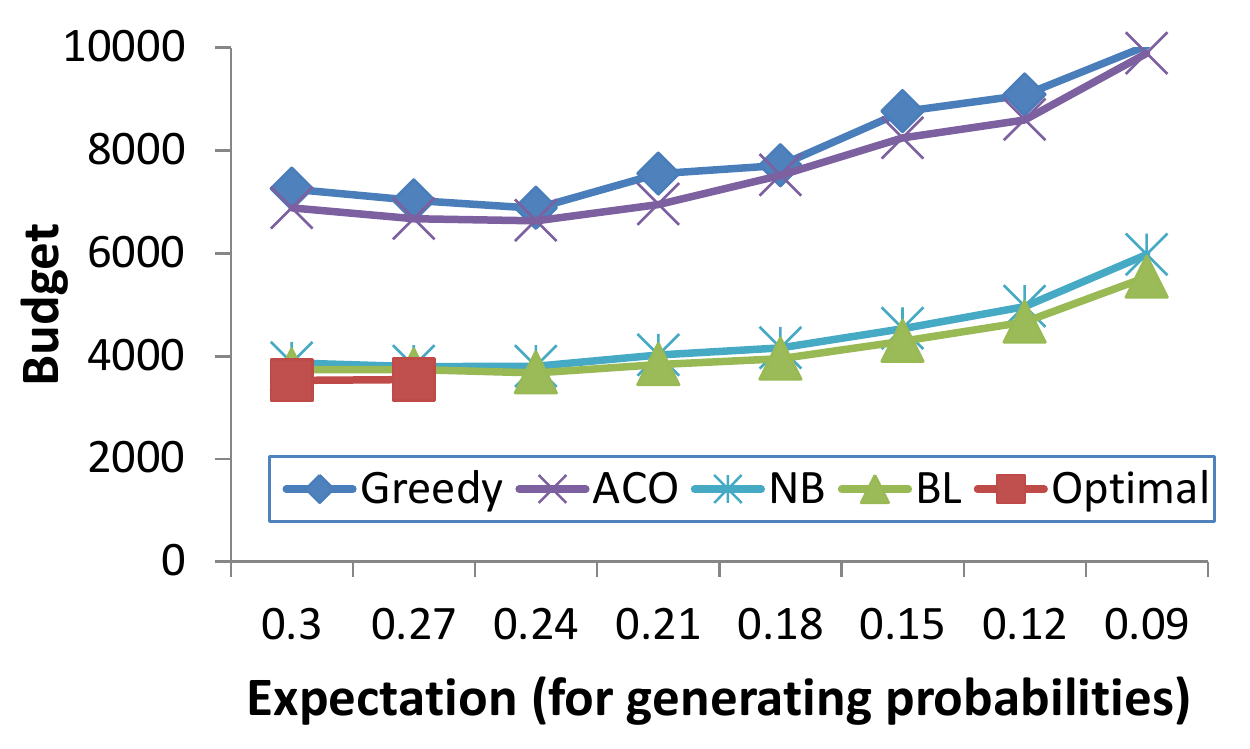}\label{subfig:exp_budget}}\hfill
\subfloat[{\small Runing time.}]
{\includegraphics[width=0.5\columnwidth]{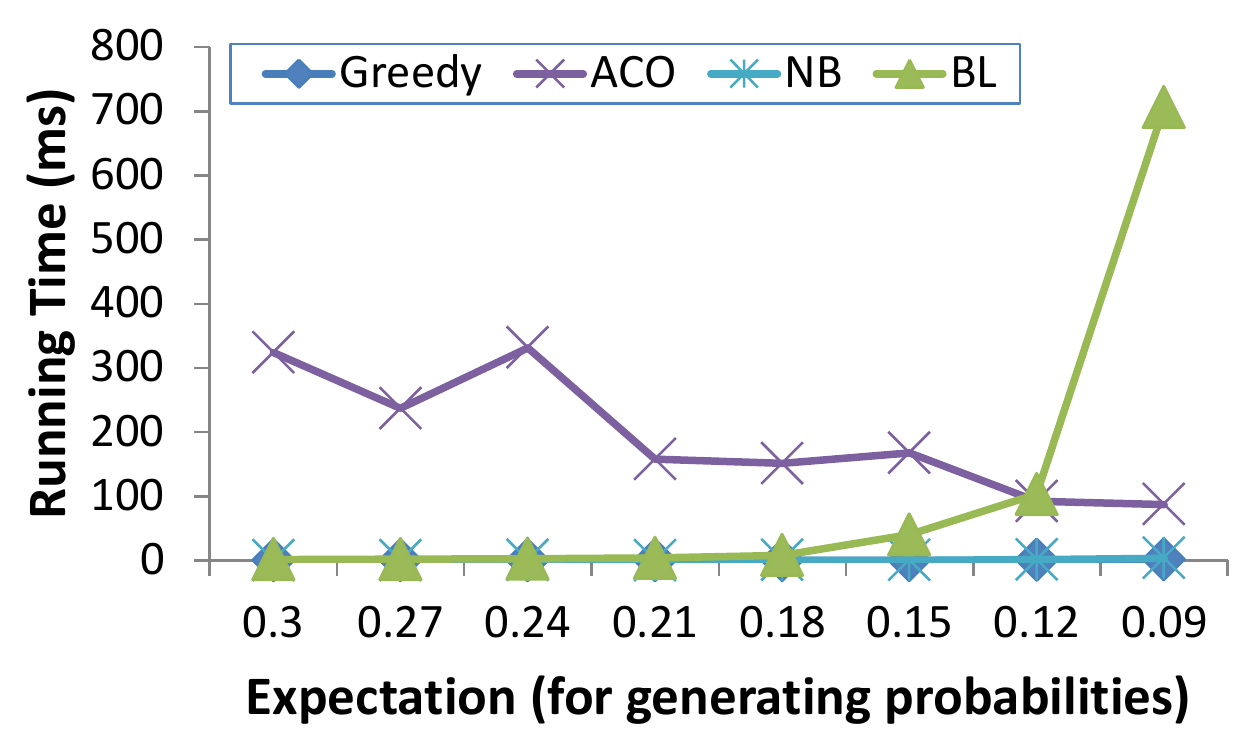}\label{subfig:exp_time}}
\vspace{-5pt}
\caption{\small Varying the expectation for generating probabilities.}\label{fig:exp}
\vspace{-8pt}
\end{figure}
As Figure~\ref{fig:prob} shows, when we decrease the expectation, which results in smaller values of probabilities, the minimal required budget increases. 
Again, BL and NB find near-optimal solutions that are statistically significantly better than Greedy and ACO. The optimal algorithm required much longer running time, and it is thus omitted from Figure~\ref{subfig:exp_time}. However,  BL demonstrated an exponential running time behavior when we decreased the expectation (for generating the probabilities). We thus conclude that NB clearly is the winner, since it runs very fast even with small probabilities and a high $p_{succ}$, but still finds near-optimal solutions. 

\begin{figure}[htbp!]\centering
\vspace{-10pt}
\subfloat[{\small Minimal required budegt.}]
{\includegraphics[width=0.5\columnwidth]{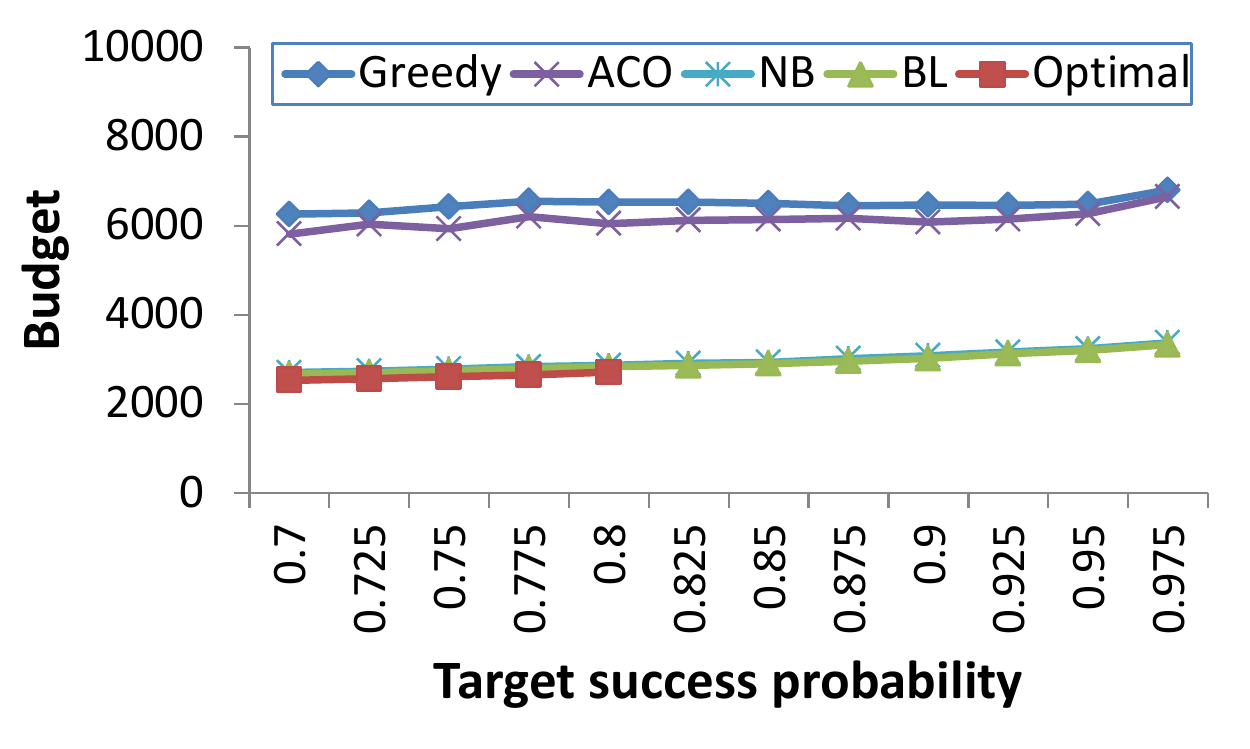}\label{subfig:prob_budget_sw}}\hfill
\subfloat[{\small Runing time.}]
{\includegraphics[width=0.5\columnwidth]{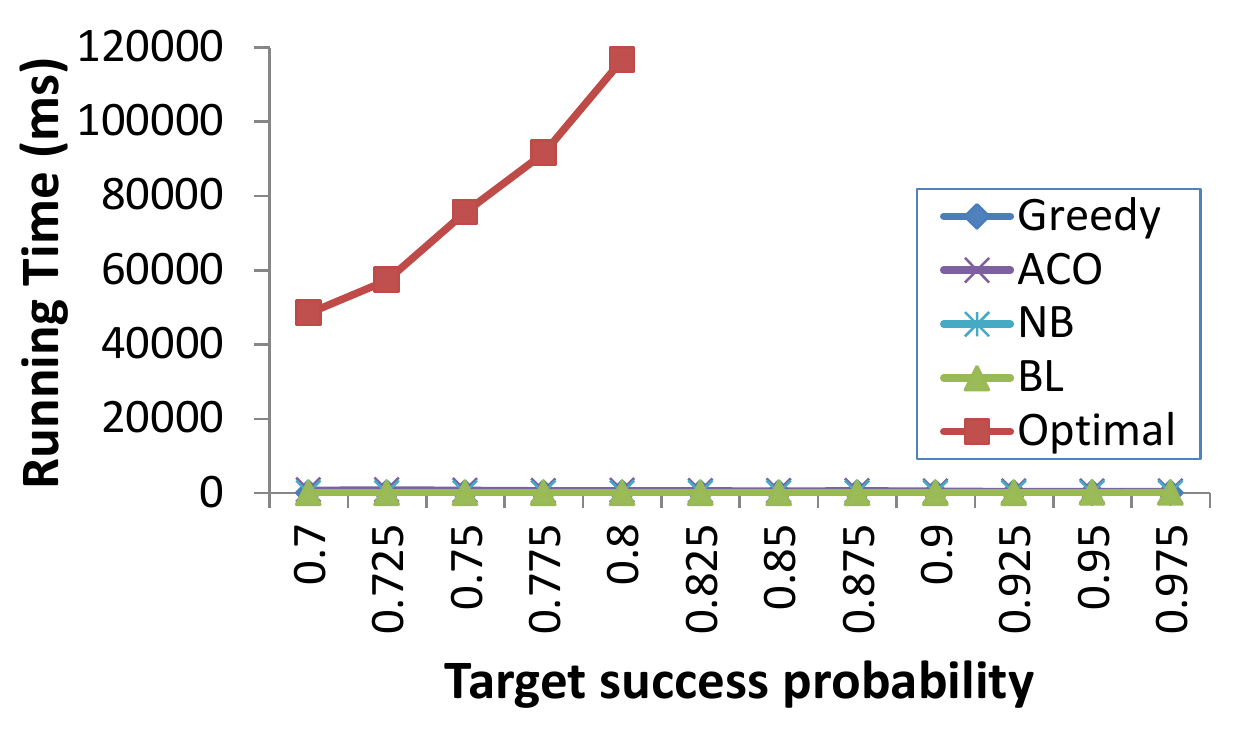}\label{subfig:prob_time_sw}}
\vspace{-5pt}
\caption{\small Varying the target success probability, $p_{succ}$, for small-world graphs.}\label{fig:prob_sw}
\vspace{-8pt}
\end{figure}

Finally, we conducted experiments on synthetic, small-world graphs with $25,000$ vertices.
Each vertex was connected to its $6$ nearest neighbors and edges in the graphs were randomly rewired to different vertices with a probability of $0.09$. The traveling cost on each edge was chosen uniformly between $40$-$1040$ so that the average edge cost will be the same as in the real graph structure setting, and the rest of the parameters were set exactly as in the real graph structure setting. The results are depicted in Figure~\ref{fig:prob_sw}, where each point is the average over $40$ randomly generated small-world graphs.
As Figure~\ref{fig:prob_sw} shows, the performance of the heuristics is quite similar to the performance with the real graph structure.
 
\section{Conclusions and Future Work}
This paper considers probabilistic physical search on graphs. We show a connection between \emph{Max-Probability} and the \emph{Deadline-Tsp} problems, which enables the $O(\log n)$ approximation for \emph{Max-Probability} with probabilities that are not too small. We believe that this connection can lead to future cross-fertilization between probabilistic physical search problems and other variants of the \emph{Deadline-Tsp} that have been extensively studied. 
We then provide a $5+\epsilon$ approximation, for every $\epsilon>0$, for a special case of \textit{Min-Budget}, and a hardness of approximation within a ratio of $1.003553$ for the general \textit{Min-Budget} problem. We further suggest several heuristics for practical use, and experimentally show that our no-backtrack branch-and-bound algorithm is able to find near-optimal solutions and handles even very large instances. We conjuncture that even NB will have an exponential running time when it will encounter specific graphs with many dead-ends and very small probabilities. However, it is possible that there are no such graphs that represent real problem instances. In addition, it is possible that Greedy and ACO heuristics will be able to handle such settings adequately, due to their almost constant running time. An important future direction is thus to explore the hardness landscape of our problems, in order to derive better insights as to which heuristic to use when. We also see great importance in extending the single-agent analysis to multi-agent settings. Finally, providing a tighter gap between the approximation and hardness of approximation results for the \emph{Min-Budget} problem remains an open challenge.

\bibliographystyle{plain}
\bibliography{tsp-prob}


\end{document}